# Experimental Reconstruction of Lomonosov's Discovery of Venus's Atmosphere with Antique Refractors During the 2012 Transit of Venus

Alexandre Koukarine[*], Igor Nesterenko[**], Yuri Petrunin[§], and Vladimir Shiltsev[§§]

[*] *Albany, CA 9470, USA*
[**] *Budker Institute of Nuclear Physics, Novosibirsk, 630090 RUSSIA*
[§] *Telescope Engineering Company, Inc., Golden, CO 80401 USA*
[§§] *Fermilab, Batavia, IL 60510 USA*
(corresponding author, e-mail: shiltsev@fnal.gov )

**Abstract.** In 1761, the Russian polymath Mikhail Vasilievich Lomonosov (1711-1765) discovered the atmosphere of Venus during its transit over the Sun's disc. In this paper we report on experimental reenactments of Lomonosov's discovery with antique refractors during the transit of Venus June 5-6, 2012. We conclude that Lomonosov's telescope was fully adequate to the task of detecting the arc of light around Venus off the Sun's disc during ingress or egress if proper experimental techniques as described by Lomonosov in his 1761 report are employed.

## I. INTRODUCTION

Mikhail Lomonosov submitted his article "The Appearance Of Venus On The Sun, Observed at the St. Petersburg Imperial Academy Of Sciences On May 26, 1761" (Lomonosov, 1761a - in Russian) for publication on July 4[th], 1761 (old style), and 250 copies were published by the St. Petersburg Imperial Academy of Sciences on July 17, 1761. The German translation (Lomonosov, 1761b) was made shortly after, presumably by Lomonosov himself, and 250 copies were printed in August 1761 for wide distribution abroad. Figure I presents a plate with illustrations from the German translation. A complete English translation of the paper with extensive commentaries has appeared recently (Shiltsev, 2012).

Lomonosov had performed the observation at his own estate in St. Petersburg (modern address Bolshaya Morskaya, 61), latitude 59°55′50″N, longitude 30°17′59″E, some 1.3 km South of the St. Petersburg Imperial Academy's Observatory. His home observatory was on the flat open roof of a 6m by 5m by 4m (length x width x height) building, equipped with ¾ m-high handrails (destroyed during reconstruction in mid-XIX century). His instrument was *"a 4 ½ feet long telescope with two glasses."* The original telescope used by Lomonosov is not preserved as it was among many 18[th]-century telescopes destroyed by enemy fire at the Pulkovo Observatory near St. Petersburg during WWII. From the 1761 report itself and its accompanying illustrations one can conclude that Lomonosov used an astronomical telescope of refractor design (with reversed image) featuring a two-lens achromatic objective. There were several indirect pieces of evidence suggest that it was possibly one of the early two-lens achromatic refractors made by John Dollond (the famous English optician, 1706-1761), but a direct evidence that Lomonosov used a Dollond achromat has only recently been found in a pre-World War II publication (for detailed discussion see Shiltsev, 2012).

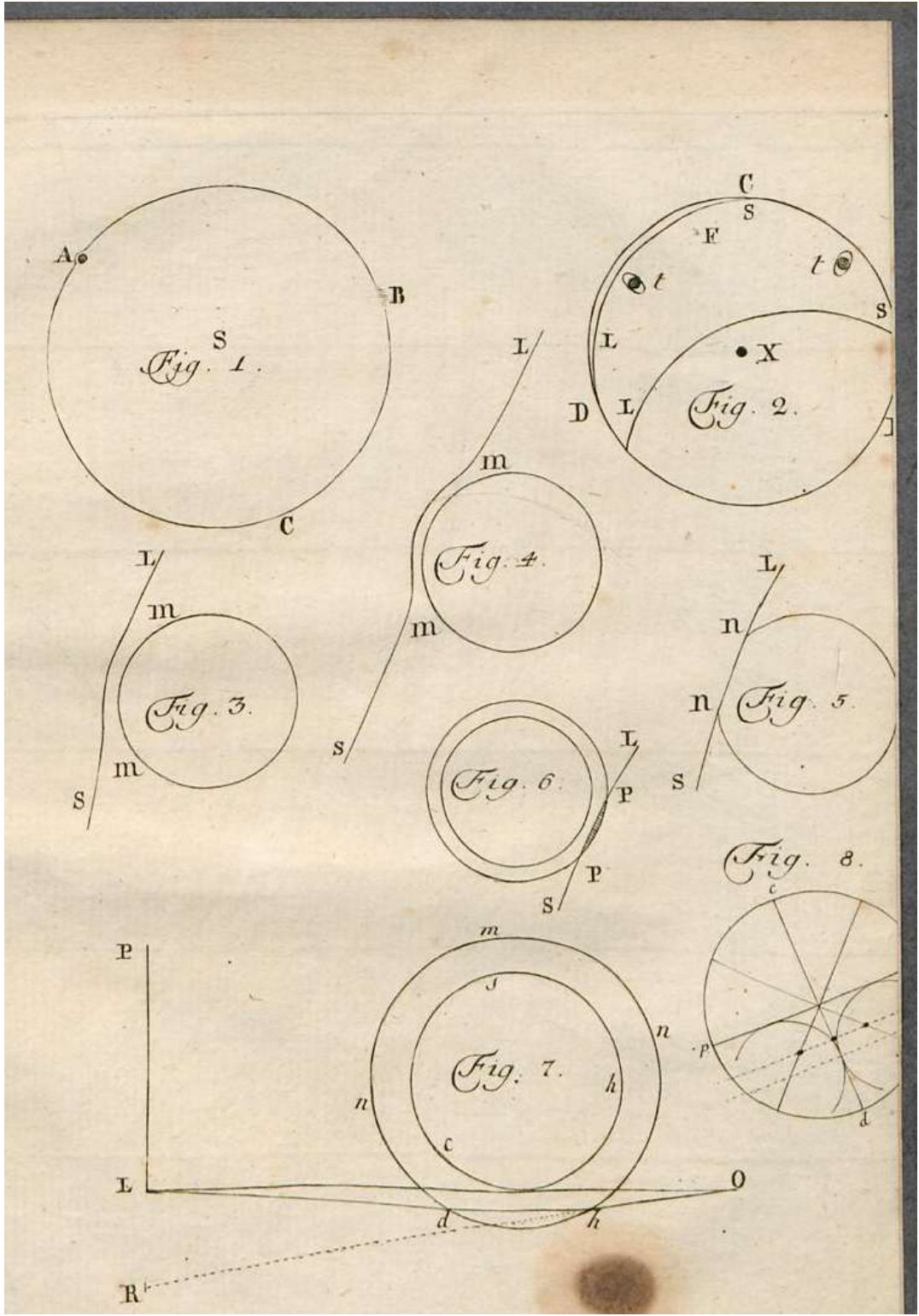

Figure I: The plate with Lomonosov's figures, from (Lomonosov, 1761b).

In his report, Lomonosov specifically mentioned that he used a very weak solar filter – "*not-so-heavily smoked glass*" - and further in the text he noted the need to give regular rest to his eyes.

Lomonosov ended the technical part of his report with "*...from these observations, Mr. Councilor Lomonosov concludes that the planet Venus is surrounded by a significant air atmosphere,*" on the basis of three phenomena observed by him (out of the full list he presented earlier): a "blurriness" of the Sun's edge at the time of $1^{st}$ and $4^{th}$ contact (illustrated in his Fig. 1 at point *B* – see Figure I, with his physical reasoning illustrated in his Fig. 6), and the "blister" or "bulge" which lasted for a few minutes after the $3^{rd}$ contact (illustrated in his Figs. 3, 4, 5 and at the point *A* in Fig. 1 – again, refer to our Figure I, with the correct physical explanation of the effect by refraction in the atmosphere, as illustrated by Lomonosov's Fig. 7). Lomonosov's Figs.3-5 indicate that the "blister" or "bulge" at 3rd contact appeared from the beginning of egress (egress phase 1.0), when Venus was fully on the Sun's disc, to egress phase 0.9-0.94.

In addition to these three phenomena, Lomonosov reported seeing "*a hair-thin bright radiance*" close to the $2^{nd}$ contact which lasted for about a second – a phenomenon which he neither illustrated, nor used in his arguments for a Cytherean atmosphere. V. Sharonov (Sharonov, 1952) argued that the "hair-thin bright radiance" close to the $2^{nd}$ contact might also be a manifestation of the refraction of solar rays in the atmosphere of Venus. In later presentations Sharonov and Chenakal (Sharonov, 1955; Chenakal and Sharonov, 1955) undertook a detailed comparison of Lomonosov's observations with reports of other observers, who mentioned similar optical effects during the 1761 transit of Venus – e.g., S. Rumovsky, J. Chappe d'Auteroch, T. Bergman, P. Wargentin and F. Mallet (effects at ingress/egress), and S. Dunn and B. Ferner (aureole around Venus while on the Sun's disc) – and they clearly established Lomonosov's priority on the grounds of: a) his precedence in publication; b) the completeness and detail of his descriptions of the observations; c) his full understanding of the observed phenomena as important physical effects and not just optical or terrestrial atmospheric nuisances; and d), his status as the sole figure to have given a correct physical explanation of the effect.

Detailed comparison of Lomonosov's 1761 results with observations of the atmospheric effects of Venus during the transits of 1761, 1769, 1874, 1882, 2004 and 2012 will be the subject of a separate analysis, but we can mention here that: a) many of the later observations were apparently similar to Lomonosov's; and b), Lomonosov did not observe the so called "black drop" effect (see, e.g., Shaefer, 2001), which was often seen during the transits.

When the 2012 transit of Venus (TOV) was approaching, a controversy erupted over whether Lomonosov could have seen the arc of light off the Sun's disc at all – e.g., Pasachoff and Sheehan (2012) questioned his discovery, citing their experience during the ToV2004 when they had difficulties seeing such a subtle phenomenon even with instruments which were supposedly far superior to the telescopes employed in the $18^{th}$ century. In this article we report on the reconstruction of Lomonosov's observations with antique $18^{th}$ century refractors. In the second part we describe the telescopes which we used for the transit observations on June 5-6, 2012, and our attempts to carefully reproduce Lomonosov's filter and experimental techniques with these instruments likely comparable to those available to the $18^{th}$-century polymath. The results of the observations are presented in the third part. We conclude with a short discussion and a summary.

## II. TELESCOPES, FILTERS AND METHODS

Four antique refracting telescopes were procured for the experiment – see photos in Figures II-IV. Their main parameters are given in Table I.

Table I: Parameters of the antique telescopes used for the 2012 transit of Venus observations.

|  | *#1-AK* | *#2-VS* | *#3-YP* | *#4-IN* |
|---|---|---|---|---|
| Maker | Dollond, London | Dollond, London | Dollond, London | C. West, London |
| Type | 2-lens achromatic refractor | 2-lens achromatic refractor | 2-lens achromatic refractor | 2-lens achromatic refractor |
| Date | ca. last third of the 18$^{th}$ century | ca. 1800 | ca. 2$^{nd}$ half of the 18$^{th}$ century | ca. 1806-1824 |
| Total length | 55" (1400 mm) | 28.3" (718 mm) | 24" (610 mm) | 18.6" (474 mm) |
| Objective Clear Aperture Diameter | 2.5" (67 mm) | 1.6" (40 mm) | 2.25" (57mm) | 1.2" (30.5 mm) |
| Magnification | 37±3 | 23±2 | 19±1 | 37±1 |
| Solar filter type | ND M3.8 | Smoked glass | ND M3.8 | ND glass M2.6 |
| Field of view | Approx. ¾ degree | Approx 1 degree | 1.2 degree | 1.2 degree |
| Solar filter attenuation at 590 nm | 1/4,000 | 1/1,700 | 1/4,000 | 1/400 |
| ToV2012 observation location | CA (USA) | IL (USA) | CO (USA) | Novosibirsk (Russia) |

The refractors (identified by the initials of the observers) were deployed for the ToV2012 observations in California (telescope *#1-AK*), Illinois (*#2-VS*), Colorado (*#3-YP*) and Novosibirsk, Russia (*#4-IN*). As the observers in Colorado and Russia (*#3-YP* and *#4-IN*) had no luck due to weather and atmospheric conditions (in addition to the small aperture deficiencies of *#4-IN*), the discussion below will center around the equipment and observations of the ToV2012 stations in California and Illinois.

*II.1 Observations in California with telescope #1-AK:*

The optical tube assembly (OTA) of the telescope *#1-AK* – see Figure II - was equipped with the original 30mm focal length Dollond erecting eyepiece of Huygens type with two additional lenses for inversion of the image. The field of view (FOV) of the telescope was approximately 45 angular minutes. The telescope exit pupil was about 1.7mm, i.e., the size of the exit pupil was near the optimum for the resolution of the human eye at middle age (Maksutov, 1946). Interferometric analysis of the telescope objective is shown in Figure III. It was taken with a 150mm ZYGO type interferometer at a wavelength of 546nm (green laser) with a precision sphere as the reference element. The test results are as follows (with reference to the diffraction limited optical parameters in parenthesis): peak-to value wave front error of 1/3.9 wavelength (1/4 or less); rms error of the wave front of 1/21 waves (1/14 or less); and a Strehl ratio of 0.916 (0.8 or more). From this one can conclude that the optics of this telescope made almost two and half centuries ago are of very good quality even by today's standards.

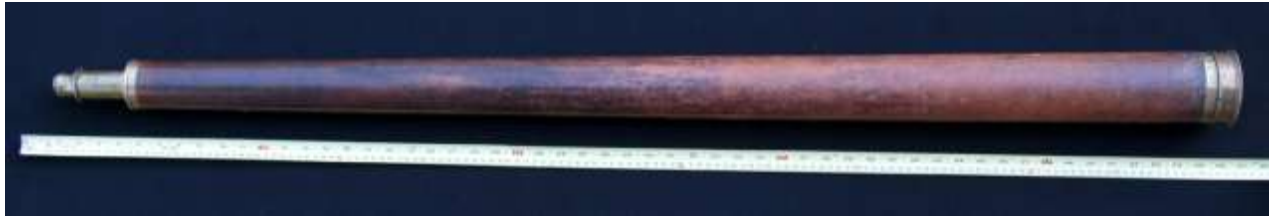

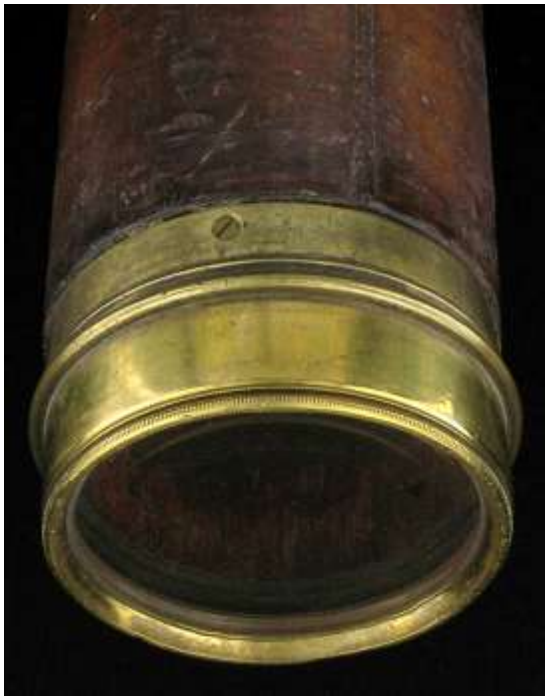 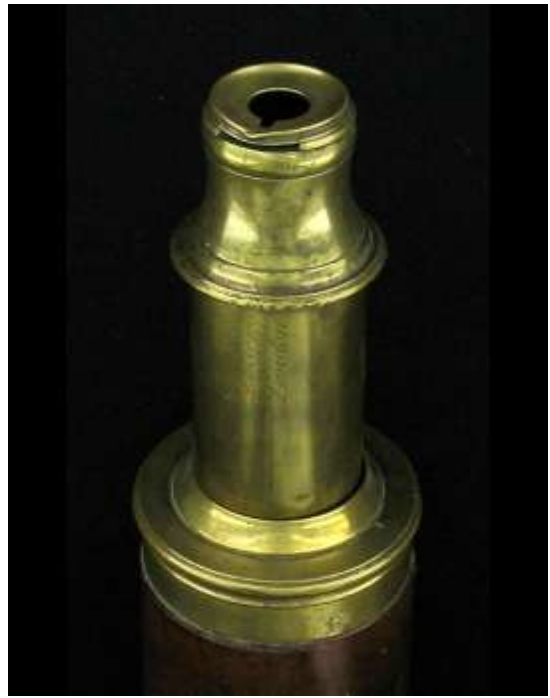

Figure II: Two lens achromatic refractor *#1-AK* by *Dollond* ca. last third of the 18$^{th}$ century – a) (top) general view; b) the objective end; c) the eyepiece end.

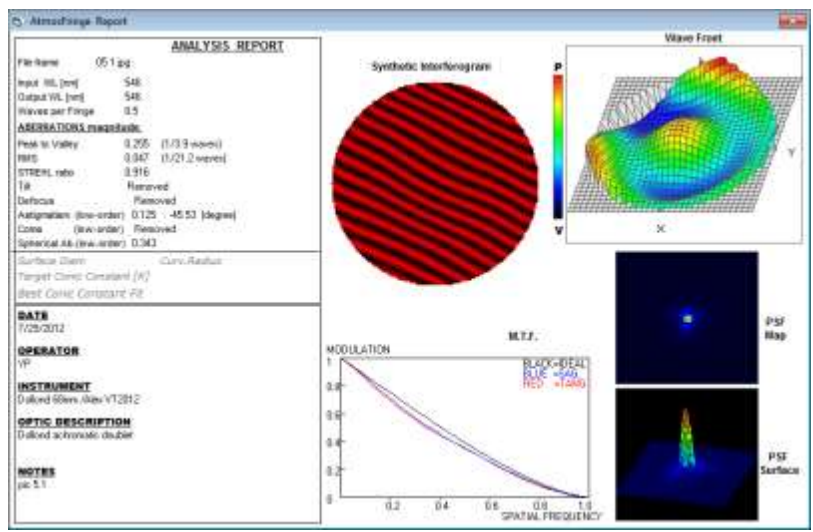

Figure III: Report of the interferometric analysis of the objective of refractor *#1-AK*.

The observer also found that the chromatic aberration of the instrument is very well compensated: the color fringe effect was noticeable only at the edge of the field of view (at approximately ¾ from the center of the optical axis). The aberration free "sweet spot," however, was shifted by about 15 minutes from the center to the right side. This minor defect in the optics was to the advantage of the observer, as he was able to keep most of the Sun's disc out of the direct view, with Venus in the middle of it. The views of sunspots and the solar surface around them were outstanding in detail and clarity on the day of the transit and during the few days prior to its occurrence.

The telescope was mounted on a generic sturdy aluminum tripod using 2 clamping rings and a generic photo tripod base, which allowed smooth altazimuth motion. The tripod's geared elevator helped to adjust the height of the eyepiece above the ground. The additional eyepiece's mass (see below) was counterbalanced with a weight mounted on the OTA close to the objective end.

Proper solar light reduction was achieved by covering the main aperture (objective) with Baader photographic AstroSolar film, with a density of M3.8, providing an attenuation of the Sun's light by some 6,000 times. The filter's spectral transmission – presented in Figure IV - was measured in the Observatory of the Novosibirsk State University (Novosibirsk, Russia). With the low-density front filter was used an Orion variable density (Moon) filter mounted in a custom adapter at the eyepiece end directly in front of the observer's eye – see Figure V. Such a setup made possible very fine adjustment of the overall attenuation in image brightness by the operator in real-time.

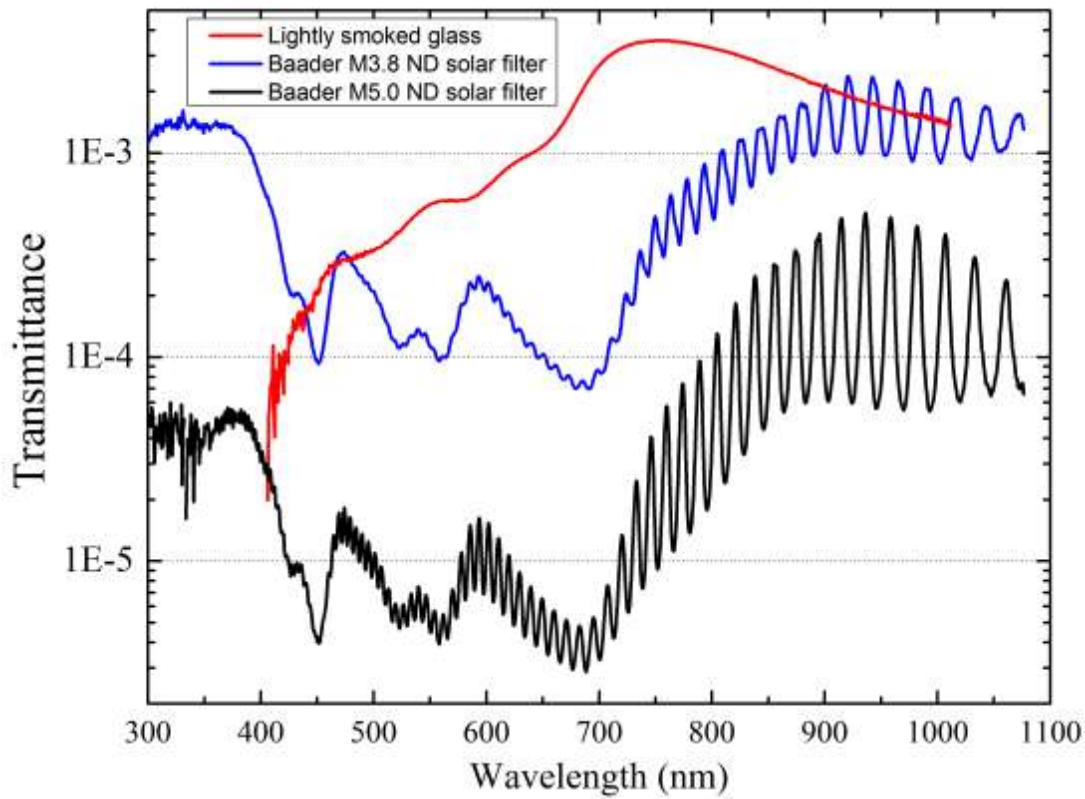

Figure IV: Transmittance spectra of solar filters: (blue) the Baader AstroSolar Photo Film ND M3.8 filter used at the objective of the *#1-AK* refractor; (red) lightly smoked glass used at the eyepiece of the *#2-VS* refractor (filter #3 – see in the text); (black) for comparison - the Baader AstroSolar Safety Film ND M5.0 recommended for safe observations.

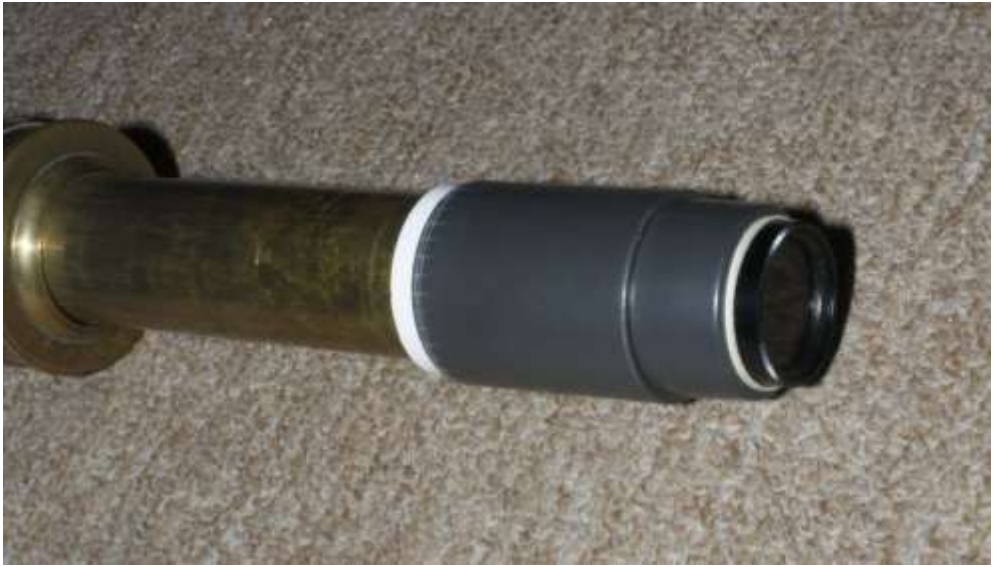

Figure V: Variable density (Moon) filter adapter used with refractor *#1-AK*.

Despite using a weak solar filter, such as the Baader AstroSolar Photo Film filter which has an attenuation coefficient 16 times smaller than that of the industry standard visual M5.0 filter, there were still concerns that it might be too dense for the task of detecting the light refracted by the atmosphere of Venus. To address this issue, we took the precaution of submitting the observer's eyes to a period of prolonged dark adaptation – a technique often used by DSO (deep sky object) observers now, as it was in William Herschel's day. It greatly helps to increase the sensitivity of the eyes to weak light conditions.

About 1 hour prior to the transit, the observer commenced to wear an opaque eye patch on his right eye under sunglasses. During the entire transit observation he had his head covered with a long black fleece hood ("balaclava helmet" or ski mask), taking care to keep its front opening tight around the eyepiece tube most of the time in order to minimize the possibility of exposure to stray light reducing his dark adaptation. He had been opening the variable density (Moon) filter completely only briefly for just 2-3 seconds once every 20-30 seconds, dimming it back after that, or changing the eye for comparison of visibility, and covering the right eye under the eye patch again.

As the Sun was quite high (around 60 degrees above the horizon), the optimal position for observation was to lie on a ground mat and use the tripod's geared elevator to adjust the eyepiece distance from the eye as necessary. Such a relaxed posture aided the observer in maintaining attention to observational detail, and delayed the onset of fatigue.

*II.2 Observations in Illinois with telescope #2-VS:*

The FOV of the telescope – pictured in Figure VI - was about 1 degree (about twice the Sun's angular size). The observer was quite impressed with the high quality of the telescope, as revealed by the sharpness of the images of the Sun, sunspots and Venus, in almost the entire FOV. No significant deterioration in sharpness was observable within ± half of the diameter of the FOV. Minimal chromatic aberration was just discernible only at the edge of the FOV, and the observer always tried to keep the north of the Sun (with Venus near it) close to the center of the FOV.

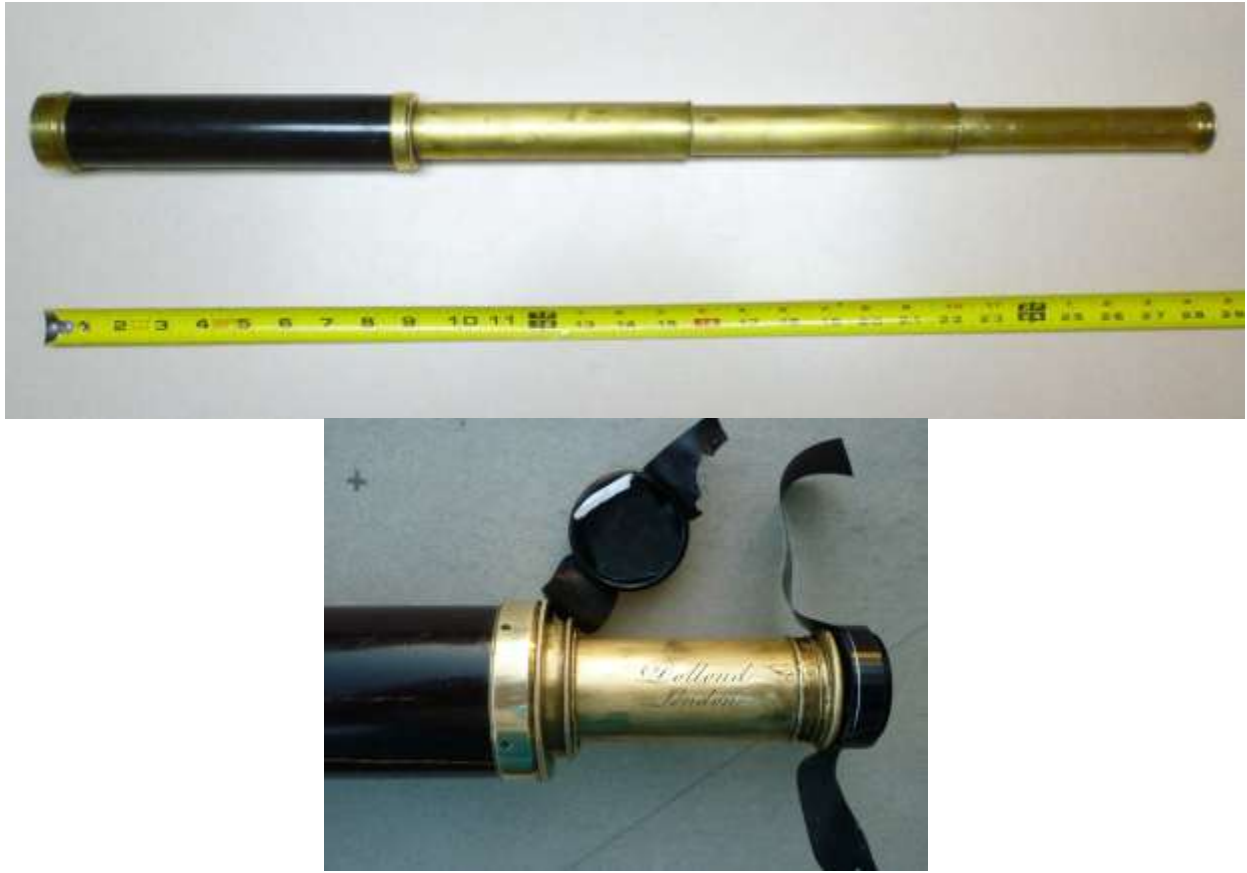

Figure VI: Two lens achromatic refractor *#2-VS* by *Dollond* ca. end of the 18$^{th}$ century-early 19$^{th}$ century – a) general view; b) eyepiece end with two smoked glass filters.

Four different filters had been prepared and tested for the ToV2012 observations with *#2-VS*: 1) a Baader AstroSolar Safe Film ND M5.0 filter (attenuation of about 100,000 in visible light); 2) a Baader Astrosolar Photo Film ND M3.8 filter (attenuation of about 6,000); 3) a hand-made smoked glass filter installed inside "15% Hirsch ND" moon filter cell; and 4), another hand-made smoked glass filter set inside a "30% Hirsch ND" moon filter cell. All the ToV2012 observations with refractor *#2-VS* have been made with **filter #3** ("15% Hirsch ND") taped right to the eyepiece, in front of the observer's eye – see Figure VI b). Figure IV shows the spectral transmission of the filter as measured in the Observatory of the Novosibirsk State University (Novosibirsk, Russia).

The choice of the filter was driven by the following considerations: a) both Baader filters appeared to have too strong an attenuation; b) the image of the Sun looked "whitish" in them; c) Lomonosov in 1761 did not have such filters, and used smoked glasses to observe the Sun. Also, Lomonosov specifically mentioned using "*not so-heavily smoked glass*" (i.e., a weak filter) and that he needed to give his eyes some rest after a short period of observation. Similar periods of rest were needed for observations with filter #3 during the ToV2012. In general, the observer used the following technique: i) place the Sun and Venus in the FOV; ii) shut the observing eye and give it a rest for some 10 sec; iii) open the eye, look at Venus and try to see the aureole; iv) continue the observation until the eye adjusts to the brightness of the Sun and sunspots begin to be visible; v) repeat the procedure (go to step ii). When the eye is first opened, the brightness of the Sun is tremendous (sunspots cannot be seen for a few seconds) but tolerable, and at the same time some finer lower intensity details can be detected. After the adaptation of the eye to the solar image (step iv) sunspots and other high intensity details were easily seen, and the only concern was to avoid keeping the very bright yellowish solar image in the eye for too long. Making preparatory runs before the beginning of the ToV2012, the observer did not feel very comfortable with filter #4 (it later was found to be 2-3 times weaker than filter #3 in the visible spectrum), though after the adaptation of the eye, he could view the Sun without serious problem, but the intense brightness at the very first moments (onto the rested eye) was worrisome. Filter #3 was therefore chosen as the most appropriate, and was taped to the eyepiece and employed for the ToV2012 observations with refractor *#2-VS*.

The telescope *#2-VS* was mounted on a metal-wood tripod. A special tent was set up to protect the telescope from direct sun and wind, in particular to avoid additional exposure of the eye to sunlight and distraction from the general public gathered to observe the ToV through other instruments. The observer was lying on the ground inside the tent, well covered by linen set around the telescope. A wrist watch and a large dial alarm clock placed in the tent were set up within a second to precise local time.

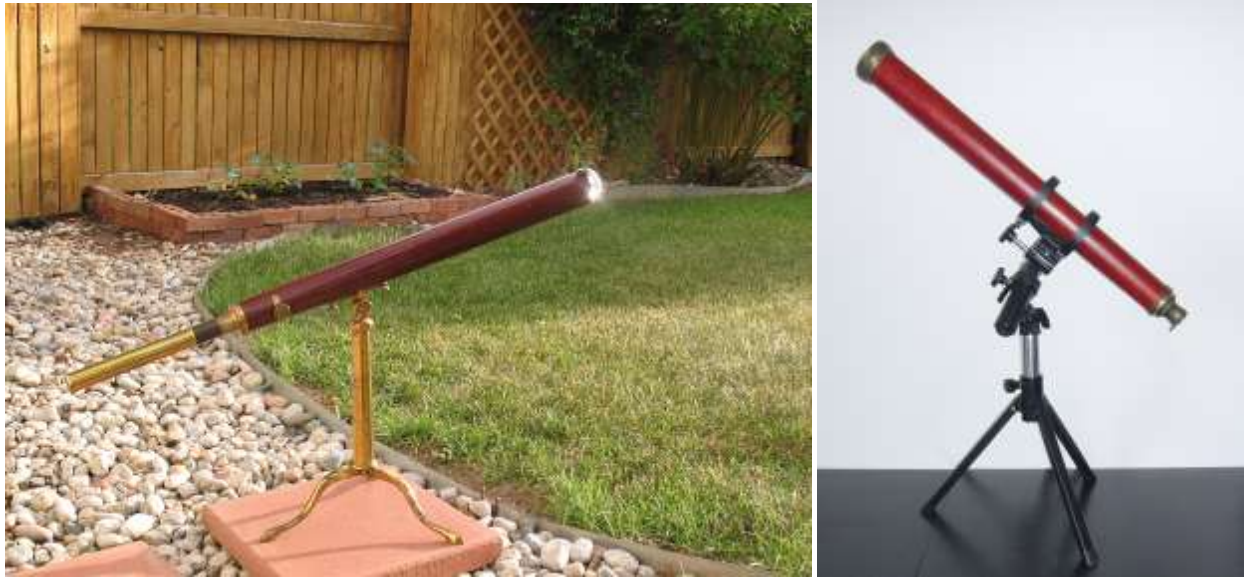

Figure VII: a) refractor *#3-YP* by *Dollond* ca. the second half of the 18th century; b) refractor *#4-IN* by *C. West, London* ca. 1st-half of the 19th century on a modern mount.

## III. THE DAY OF OBSERVATIONS AND RESULTS

Both ToV2012 observations with refractors #1-AK and #2-VS took place on June 5th, 2012 and both observers were able to see **only the ingress** of the planet on the Sun's disc. A brief summary of the transit conditions at both locations is given in Table II. More detailed accounts and results of the observations are presented below, separately for each observer.

Table II: Conditions at the observation stations on the day of the ToV2012 (June 5, 2012)

|  | *Lick Observatory, CA* | *Batavia, IL* |
|---|---|---|
|  | *#1-AK* | *#2-VS* |
|  | Latitude 37°20′52″ N | Latitude 41°51′00″ N |

|  | Longitude 121°37′23″ W | Longitude 88°18′45″ W |
|---|---|---|
| Time of first contact | 15:06:30 UT-7 | 17:05:08 UT-5 |
| Time of second contact | 15:23:58 UT-7 | 17:22:39 UT-5 |
| Height of the Sun | 58 deg. | 34 deg. |
| Azimuth of the Sun | 250 deg. | 90 deg. |
| Air temperature | 44 deg F (7 deg C) | 68 deg F (20 deg C) |
| Air pressure | 28.5 in | 30.00 in |
| Wind | N-W, 13 mph (felt like 2 mph behind the mountain ridge) | N-E, 14 mph |
| Visibility | (excellent) | 10 mi. |
| Clouds | variable | none |
| Elevation | 4000 ft. | 696 ft |

*III.1 Observations in California with telescope #1-AK:*

In order to minimize the possibility of cloud cover in the local weather forecast for the day of the transit (June 5, 2012), and to minimize hot air instability effects possible at sea level in the San Francisco Bay Area, it was decided to set up the instrument on the territory of the Lick Observatory, which is owned and operated by the University of California and situated on the summit of Mount Hamilton, San Jose (California, USA). The exact location coordinates are Latitude 37.346534, Longitude -121.623728, and Elevation 4000 ft. This remote location allowed the observer to avoid the overcrowding anticipated during the Observatory's public observation event, as well as provided a better chance of protection from the usual afternoon westerly wind. The actual weather conditions were not really favourable to observers at first glance. Many clusters of low cumulus cloud covered the entire sky, and were moving slowly in a south-easterly direction. There were promising gaps between them, however, and the sky behind looked crystal clear, free of any cirrus layers. The atmospheric transparency was outstanding on the day of the transit following the rain showers the day before. The observer was able to see clearly the Sierra Nevada Mountain ridges behind the San Joaquin Valley, which are more than 100 miles away to the east.

Due to the variable cloud cover, the observer enjoyed only 4 windows of excellent viewing. He did not register the time of events, and did not make any drawings at the time of the observations, as he was concentrating on preserving his darkness adaptation, maintaining the optimum sensitivity of his eyes, and memorizing everything he saw. The first prolonged time window provided the view of the first contact down to approximately 1/4 of the ingress of Venus' disk (the ingress phase 0.25). There were no unusual visual effects noted by both eyes at this stage.

The second time window was rather short - about 2 minutes - and provided the opportunity to see the atmosphere of Venus in direct view as an aureole at the phase of ingress of approximately 0.7 to 0.75. It looked like a hair-thin smooth arc starting from the North side of the disk of Venus and extending by a bit more than half-way to the other side. It was very thin but clearly visible at the West (right) end of the arc. The observer's drawing is presented in Figure VIII a). The time window of the observation was in fact too short to estimate the overall dynamics precisely, but the observer did perceive the partial arc as slowly extending farther to the right side with the phase of ingress growing. The left, less prepared eye also perceived the arc; it was not visible in the direct view, however, but only when the observer used averted vision. He also tried to move Venus around in the FOV, looking for changes in arc visibility. This did not change the shape of the arc at all, except when hitting the very edge of the FOV and blending with the chromatic diffusion halo (certainly a testament to the quality of the Dollond achromat's optics).

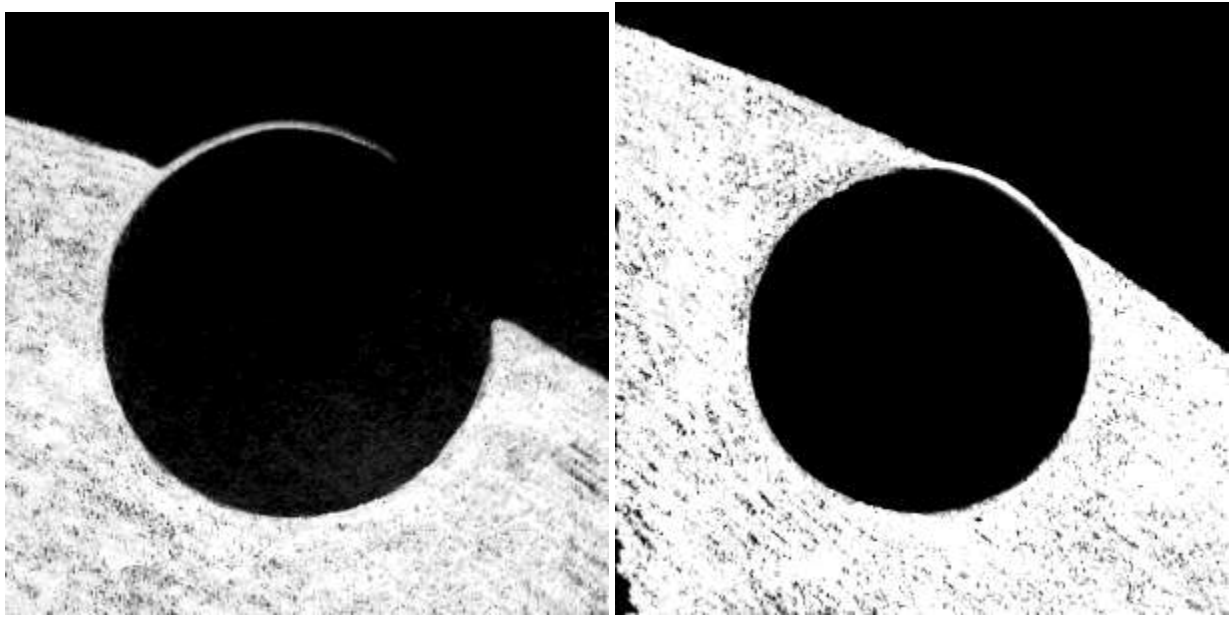

Figure VIII: Observations with refractor *#1-AK* during the ingress of TOV2012: a)the arc of light slightly over the middle of the black gap; b) the bulge on the edge of the Sun,

The third opportunity arrived just in time to observe the bulge on the edge of the Sun as Venus' disk sank completely. It started at very nearly ingress phase 0.9 and was visible through the thin edge of a passing cloud; at first the observer did not see any light from the arc at all, just a black gap. When the cloud went off completely the observer thought he has missed the arc, as the Sun's edge looked evident above it, however when he opened the filter to the maximum again he noticed that the edge above in fact had a tiny but wide irregularity in the smooth progress of the Sun's edge curve. The bulge was the atmosphere of Venus revealing itself again – see Figure VIII b). It had disappeared at an unremarked moment, probably when the observer tried to change to the other eye. After returning back to the right eye the bulge could no longer be detected.

Observation of ingress continued for another minute or so, in an attempt to see the so called "black drop" effect or similar phenomena. However, the only thing observed was just the bright gap between the edges of the Sun and Venus which grew without any visible peculiarities.

*III.2 Observations in Illinois with telescope #2-VS:*

Observations were made from the backyard of a private estate in Batavia, IL (USA), latitude 41°51′00″.0 N, longitude 88°18′45″.4 W, elevation 696 ft (212 m). The sky was absolutely clear throughout the stages of ingress and remained so till sunset. The solar altitude in degrees above the horizon and the azimuth (the compass direction in degrees, measured Westward from the South) are given in the Table II. The temperature at the time of ingress was about 68°F (20 °C), with a slight wind.

The observer reported that it was hard to time exactly the moment of 1st contact - he was able to fully appreciate seeing the leading edge of the disc of Venus on the Sun at 17:05:05 CST (though he had doubts for some time before that moment). The ingress progressed uneventfully for the next 5 min. Around 17:11:00 a little light "whisker" appeared on the left (Northern) side of Venus expanding by less than 0.1-0.2 of its diameter off the Sun's disc – see Figure IX a). The observation was repeated again, and then with another eye. By 17:13:00 it became certain that the "whisker" was not an artifact. There was no similar "whisker" of light on the opposite (right) side of Venus, instead the solar limb looked slightly bent inside the Sun's disc.

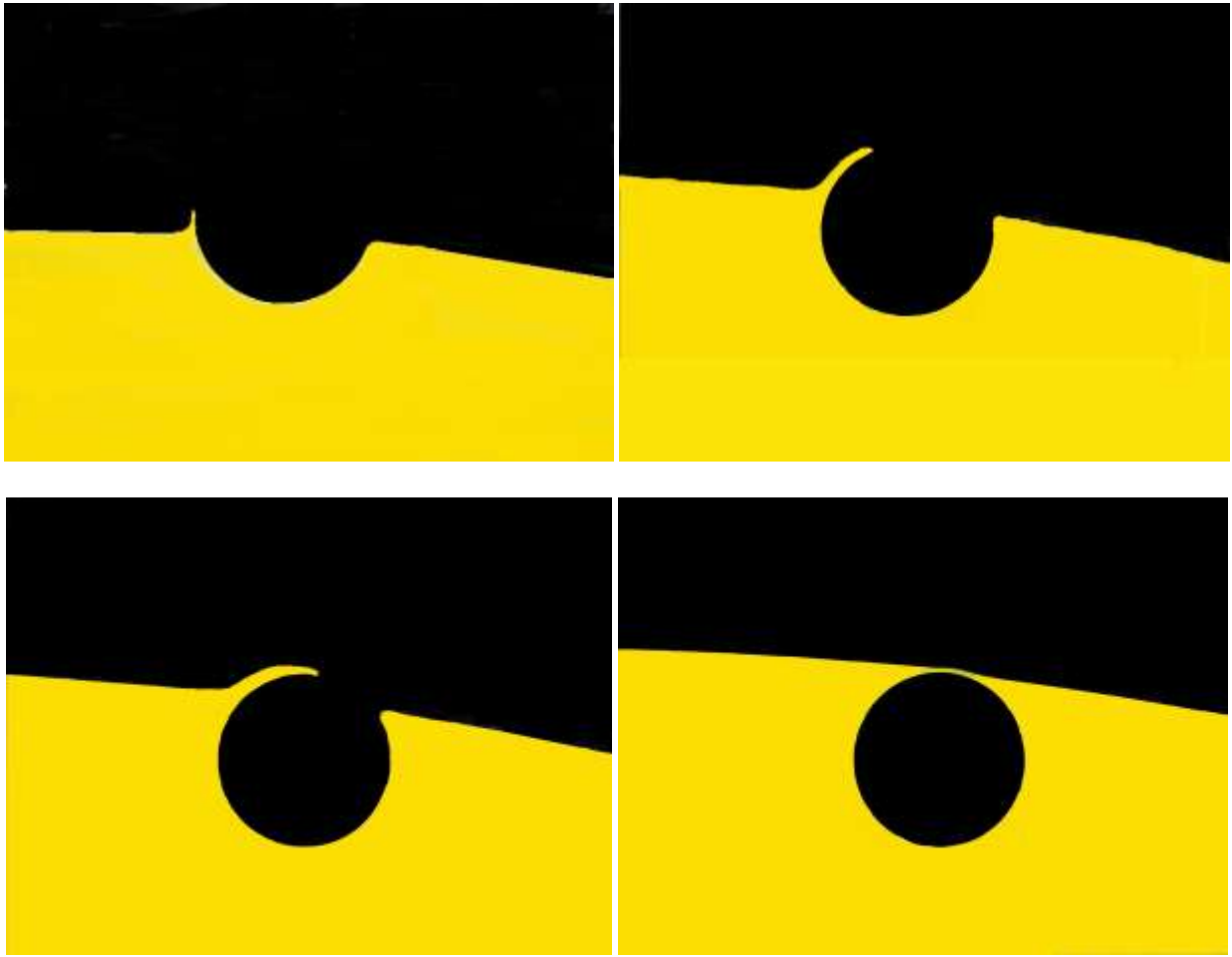

Figure IX: Venus during various stages of the ingress as observed with refractor *#2-VS* (left to right): a) at about 17:11; b) at 17:16; c) at 17:19, e) between about 17:21 and 17:22 (all times CST=UT-5).

Around 17:16:00, when Venus was about half way on the Sun, the "whisker" (partial arc of light around the planet) had started to be reliably seen and somewhat increased in length up to about 1/4-1/3 of half of the circumference of Venus' disc – see Figure IX b). The length of the "whisker" extended to about 1/2 of the circumference of Venus off the Sun by about 17:19:00 – see Figure IX c). The left and right sides of the Sun met over Venus around 17:20:15, and though the "bulge" or the full arc was not prominent, the left-right asymmetry was still discernible. Until 17:22:00 the left-arc and the right-arc were in a sort of connected position with a shady area between the disc of Venus and the darkness off the Sun as shown in Figure IX d).. Only after 17:22:00 CST did the surroundings of Venus appear to be fully left-right symmetric. Nothing particularly exciting happened over the next three minutes (e.g., no prominent "black drop effect" was seen), as Venus smoothly moved over Sun's disc and observations with the refractor #2-VS ended around 17:26:00.

To summarize, between 17:11 and 17:20:15 the observer had clearly seen the partial arc of light ("whisker" or "fang") on the left (Northern) side of Venus, and around the disc of Venus the arc had initially started small and expanded over the disc of the planet. Between 17:20:15 and 17:22:00 the observer reported definitely seeing light, though not as bright as everywhere else on the disc, which connected over the entire disc of Venus, although it was not clear whether Venus was fully on the Sun or was still partly off it. Around 17:21:00 the observer ended up in a very uncomfortable position after moving his body (he was lying down at the time), and as a result, he was forced to bring his head up (off the support) and keep it up so as to be able to continue observing. Keeping the head up all the time was hard, and required resting the neck muscles every so often, which might have compromised the consistency and quality of observations over the last 1.5 minutes of ingress. As a result, the observer was not certain whether he saw the complete "Lomonosov's arc" (bulge) in the last 105 seconds prior to 17:22:00 or not.

## IV. SUMMARY

Several conclusions can be drawn from the 2012 transit of Venus observations with antique refractors presented above:
1) Lomonosov's arc was observed with the 4.5-ft two-lens Dollond achromat refractor (*#1-AK*), an instrument similar to the one used by Mikhail Lomonsov in 1761;
2) the partial arc of light ("whisker") around the Northern part of the disc of Venus off the Sun was also observed during ingress with the "Lomonosov-like" refractor (*#1-AK*) and with another somewhat smaller 2.4-ft two-lens Dollond achromat refractor (*#2-VS*) from around the end of the 18$^{th}$ century;
3) weak solar filters and special techniques to preserve the eye's sensitivity at a maximum - similar to the ones described by Mikhail Lomonosov in his report – were of great help in making possible observations of Cytherean atmospheric effects ("Lomonosov's arc" and the "whiskers") with the 18$^{th}$ century refractors;
4) in general, we can conclude that an 18th-century observer viewing the transit of Venus in 1761 through a Dollond doublet achromatic refractor, similar to the those used by us (*#1-AK* and *#2-VS* ) and employing weak solar filters could have seen the aureole around Venus caused by the refraction of light in the atmosphere of Venus ("Lomonosov's arc").

Additionally, we should point out that neither of the two observers reported seeing any prominent "black-drop" effects with the antique 18$^{th}$-century refractors equipped with weak solar filters or noticing "blurriness" of the Sun's edge at the very time and location of the 1$^{st}$ contact (and both reported an indeterminacy in the timing of first contact). We should also emphasize the overall satisfaction of the modern observers with the superb image quality provided by the antique Dollond achromats. This study adds yet another argument in support of Lomonosov's priority in the discovery of the atmosphere of Venus, as he was the first and the only observer of the 1761 transit of Venus who realized the need of, implemented and described in his report the experimental methods (weak solar filter and increasing eyes sensitivity by regular rest) which allow successful reproduction of the aureole observations even after two and half centuries.

## ACKNOWLEDGEMENTS

The authors would like to thank many people who helped us to make this re-enactment possible: Profs. R.Z. Sagdeev, V.G. Dudnikov and G.I. Dudnikova for cooperation in arranging the observations; Profs. B.S. Elepov and I.A. Maltsev for providing copies of numerous scholarly articles on Lomonosov's scientific achievements; Profs.

R.C. Crease and R.A. Rosenfeld for helpful discussions of the details of the reconstruction and the results of the observations. The many useful comments we received from the participants of the *astronomy.ru* forum, where we posted our draft reports, are greatly appreciated. Special thanks to Prof. R.A.Rosenfeld for careful reading and commenting the manuscript.